# New water oxidation mechanism in Photosystem II resolves major experimental controversies.

Yulia Pushkar

[1] Department of Physics and Astronomy, Purdue University, West Lafayette, Indiana 47907, USA

*Correspondence to:  ypushkar@purdue.edu

## Acknowledgements

This research was supported by NSF, CHE- 2303743 (Y.P.).

**Abstract**

Light driven oxygen formation in Photosystem II protein is a fundamental process that sustains our biosphere and serves as a blue print to future clean energy solutions due to its high energy conversion efficiency. Last decade of intense research by advanced physical techniques delivered new insights on the structure and function of the $Mn_4CaO_5$ cluster – a center of the oxygen evolving complex (OEC). However, discrepancies in experimental observations and computational models persist impeding the understanding of the O-O bond formation and the role of the protein environment in the process. Here we show that i) assignment of the OEC unique oxygen O3 ligated by histidine (His337) via dynamic H-bond as a slow exchanging substrate and ii) its coupling with O6 oxygen generated at Mn1 in the $S_2$ to $S_3$ transition - give the O-O bond formation mechanism most consistent with all currently available experimental data. Proposal shows how protein environment can steer the O-O bond formation by charge control via H-bond and open coordination of Mn1. Obtained O3-O6 peroxide is at lower energy than peroxides in the most studied O5-O6 bond formation pathway. His337 appears to be similar to distal His in globins used for management of the $O_2$ and $H_2O_2$ intermediates. The new mechanism breaks the prior impasse and will undoubtedly invigorate future detailed studies uncovering its further details.

## Introduction

Photosystem II (PS II) is a membrane-bound protein–pigment complex found in plants, algae, and cyanobacteria that catalyzes the light-driven oxidation of water to molecular oxygen, thereby sustaining the aerobic life on Earth. At the heart of the light-induced water oxidation is the Oxygen-Evolving Complex (OEC), a catalytic $Mn_4CaO_5$ cluster embedded within the protein environment.[1-5] The cluster has intricate ligand environment with total of six carboxylates bridging it's metal ions: three are bridging Mn-Mn pairs and three are bridging the Mn-Ca pairs.[4, 6], one His332 directly coordinates Mn1 via Mn-N bond and one His337 provides the hydrogen bond to a bridging oxygen atom with a number O3, **Figure 1**. Ca and Mn4 are also coordinated with $OH_x$ fragments, which are most probably both waters on $Ca^{2+}$ and can be a combination of waters and -OH for Mn4. Protonation states of all bridging oxygens are not known with certainty and can depend on the S-state of the Kok cycle, **Figure 1A**. Kok cycle serves as a framework for understanding the overall OEC function as it explains how 4 light-induced electron-transfer events couple to the 4 electrons ($2H_2O = 4H^+ + O_2 + 4e^-$) water oxidation reaction. Here we will focus on more oxidized $S_2$ and $S_3$ states and pathways of O-O bond formation in the $S_3$ to $S_0$ transition, while $S_0$ and $S_1$ states are more reduced. The $S_2$ state with a spin S=1/2 configuration provided a wealth of coordination and electronic structure information via EPR measured hyperfines, such as $^1H$; $^{14/15}N$, $^{55}Mn$ and $^{17}O$.[7, 8] While we have an abundance of structural information[4] and information about localization of the spin-density (at least in the $S_2$-state), we still do not have the confirmation on the mechanism of the O-O bond formation as it remains a topic of debate.[9-11] This difficulty stems in part from the fact that OEC has many oxygen atoms and its environment also features water molecules which can potentially participate in O-O bond formation via water nucleophilic attack mechanism.

The important analysis which complements the OEC structural studies is the study of the oxygen substrate exchange by time-resolved isotope ratio membrane-inlet mass spectrometry in combination with select spectroscopic techniques.[7, 12-15] The most significant substrate water exchange results can be summarized as follows. In the $S_0$-state, only one substrate water is observed to exchange at a rate of $10\ s^{-1}$. In $S_1$, the substrate water exchanges at a rate of only $0.02\ s^{-1}$, and in $S_2$ the rate increases to $2\ s^{-1}$. In $S_2$, a second substrate water is observed to exchange almost as fast as if it was not bound. In $S_3$, there is a slow substrate water exchange at a rate of $2\ s^{-1}$, and a fast exchange at $40\ s^{-1}$. The two most interesting of these observations, both quite unexpected, are that the slow exchange is actually faster in $S_2$ than in $S_1$, and that the slow exchange is the same in $S_2$ and $S_3$. In addition, the $Ca^{2+}$ cofactor substitution with $Sr^{2+}$, either by chemical exchange or biosynthetic incorporation, enhances the rate of $W_S$ exchange but leaves $W_f$ unchanged. [7, 12, 15, 16] This suggests that $W_S$ is either associated with the $Ca^{2+}$ site such as been a bridge between the Mn-Ca or steps of $W_S$ exchange are affected by at least temporary coordination between the Mn and Ca.

In this manuscript we deliver a new assignment of the fast and slow exchanging oxygens and mechanism of O-O bond formation which follows from this new assignment. Outlined mechanism is currently the one most consistent with all available experimental data and resolves controversies of previous mechanisms which currently impede the progress in analysis of OEC and design of its biomimetic assemblies.

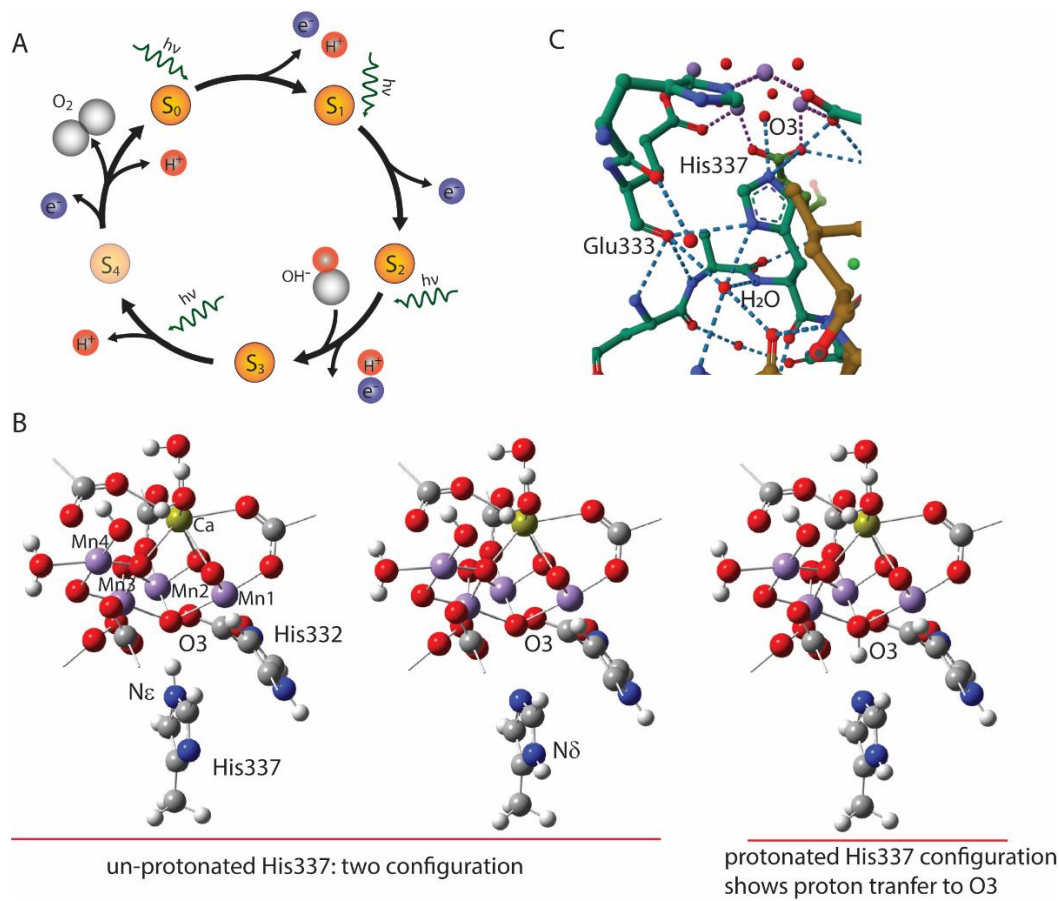

**Figure 1.** A) Scheme for the Kok cycle showing light-induced electron transfer events and definition of the S-state intermediates. B) Model of the Oxygen Evolving Complex of the Photosystem II in the $S_3$-state. Four Mn and one $Ca^{2+}$ ions are shown with six bridging carboxylate ligands and two histidine. Side chains are truncated to provide clear view. Interaction between the His337 and O3 oxygen are analyzed as having three possible configurations where i) O3 is protonated (O3H) as this configuration is relevant for slow substrate exchange); ii) in H-bonding with N-H group; and iii) lacks H-bond – a state which lowers the energy formation of O3-O6 peroxide. C) visualization (from 8IR5) of the His337- $N\delta$ nitrogen environment. $N\delta$ can form H-bond with backbone of Glu333 and conserved water molecule which is involved with extended H-bonding network.

## Results and Discussion:

*Recent assignment of O5 as Ws faces several contradictions.*

The most recent detailed study performed parallel substrate water exchange experiments in the $S_1$ state of native Ca-PSII and biosynthetically substituted Sr-PSII employing Time-Resolved Membrane Inlet Mass Spectrometry (TR-MIMS) and a Time-Resolved $^{17}O$-Electron-electron Double resonance detected NMR (TR-$^{17}O$-EDNMR) approach.[7] Authors interpreted results as a correlation of oxygen bridge between the Mn3 and Mn4 ion (O5) exchange with the same rate as a rate of exchange as the slowly exchanging substrate water (WS) in the TR-MIMS experiments and ii) that the exchange rates of O5 and Ws are both enhanced by $Ca^{2+} \rightarrow Sr^{2+}$ substitution in a similar manner. Conclusion was made that only O5 fulfills all criteria for being $W_S$.[7] Some deliberations were also presented as to exclude water ligand on Mn4 as been not suitable for Ws assignment.

However, the O5 is a bridging oxygen in a di-μ-oxo Mn4-Mn3 unit which is not protonated. Such bridging oxygens between high valent Mn ions exchange extremely slowly ($<10^{-2}$ s$^{-1}$ for $Mn_2$(III,IV) and $<10^{-6}$ s$^{-1}$ for $Mn_2$(IV,IV)). In addition this bridging oxygen been deprotonated means that its binding energy to a metal ion should increase and thus decrease its exchange rate strongly. This first contradiction has been explained away with detailed DFT calculations, however, the feature of the back electron transfer from the $S_3$ state to Tyrz was required to show that exchange is still possible in the $S_3$ state.[17] Calculations also featured water coordination to otherwise 5-coordinated Mn1 with subsequent protonation of the O5 and transition to closed cubane conformation necessary in all $S_1$, $S_2$ and $S_3$ state exchange reaction schemes.[17] Guo et al. repeated some of these calculations later essentially reproducing main results and emphasizing ones again the requirement of open – closed cubane interconversion to model Ws water exchange at O5.[18] Currently closed cubane remains to be experimentally unconfirmed geometric state in the OEC.[19, 20]

Second contradiction is that measured hfs associated with $^{17}$O do not exceed ~8 MHz[7], which is extremely low for bridging oxygen. It is well known that in the OEC there is a significant hole delocalization onto the bridging oxygens of the di-μ-oxo Mn-Mn units such as O5. Thus, $^{17}$O hfs might reach, at least theoretically, as high as 60-50 MHz on some of their components, **Table S1**. **Table S1** shows $^{17}$O hfs computed for several DFT models of the $S_2$ state, S=1/2. Some of these models were reported earlier[20, 21], others are re-optimizations with a different protonation states, **Table S2**. Note that water coordinated to Mn4 was predicted to have low hfs (**Table S1**) in agreement with earlier analysis concluding that it is unlikely to serve as a substate oxygen. Interestingly the O3 came as the one candidate oxygen featuring "intermediate" hfs likely due to H-bonding with the His337. pKa of His is ~6, thus, at typical pH of PS II buffer solution this residue can be protonated as well as de-protonated. **Table S2** compares DFT calculation of the $S_2$ state model with protonated His337 and unprotonated His337. His337 protonation results in small ~0.1 Å elongation of the Mn-Mn distances in the Mn3-Mn2 and Mn2-Mn1 bridges. Significantly His protonation results in the changes in the H-bonding pattern between the His337 and O3 where proton of the H-bond between the O3---H-His shifts to protonate O3: O3-H---His. These results are similar to the computational analysis reported earlier.[22] For $S_3$ model with Mn1=O fragment protonation of His results in similar elongations of Mn-Mn bonds of the open cubane core by ~ 0.07-0.09 Å, **Table S2**, **Figure 1B**. O3 becomes protonated in the $S_3$ state model with protonated His337. Thus, pKa of this bridging oxygen is close to ~6 in both analyzed $S_2$ and $S_3$ states showing that its protonation can be functionally relevant.

There were some proposals that O4 bridging Mn4 and Mn3 might also be protonated[23] but such proposals were not confirmed from energetic standpoint.[24] Overall, the comprehensive analysis of the $^{17}$O hfs for the range of the $S_2$-state models in the S=1/2 state has shown that O5 does not fit the experimentally measured ~8 MHz signal. However, O3 is significantly more consistent with the properties of Ws from both EPR and substrate exchange properties as will be shown below. His337 protonation further decreased $^{17}$O hfs of the O3 in the $S_2$ state, **Table S1**. O3 is a frustrated oxygen placed in between four dissimilar positive charges: three Mn ions and one proton. Thus, O3 distances to Mn are longer than in a typical di-µ-oxo Mn-Mn unit resulting in lower hole delocalization from Mn ions into O3, lower $^{17}$O hfs and potentially lower binding energy which facilitates the O3 exchange on sub-seconds time scale. As we concluded that O3 is a more suitable oxygen center to explain the experimentally determined ~8 MHz $^{17}$O hfs splitting[7] we continue to discuss how this center can meet the substrate exchange requirements imposed by experimental observations.

*Plausibility of the O3 assignment as Ws*

The advantage O3 has over the O5 as a potential slow exchanging substrate oxygen is that it is the only bridging oxygen in the OEC which can be reliably shown to be protonated – a necessary pre-requisite for exchange. Protonation of the His337 at Nδ results in movement of the proton from its Nε to O3 (**Table S2**). We foresee the exchange to have same start as in the earlier proposed scheme for O5 exchange [17] – with the coordination of water to five coordinate Mn1. Then, instead of proton transfer to O5 and complex open to closed cubane interconversion needed for the O5 exchange, is it simply enough to transfer a proton from water coordinated to Mn1 to protonated O3 and the water ligand in the position of O3 is ready to leave and remained -OH coordinated to Mn1 can take its position ensuring sub-seconds exchange rate, **Figure 2A**.

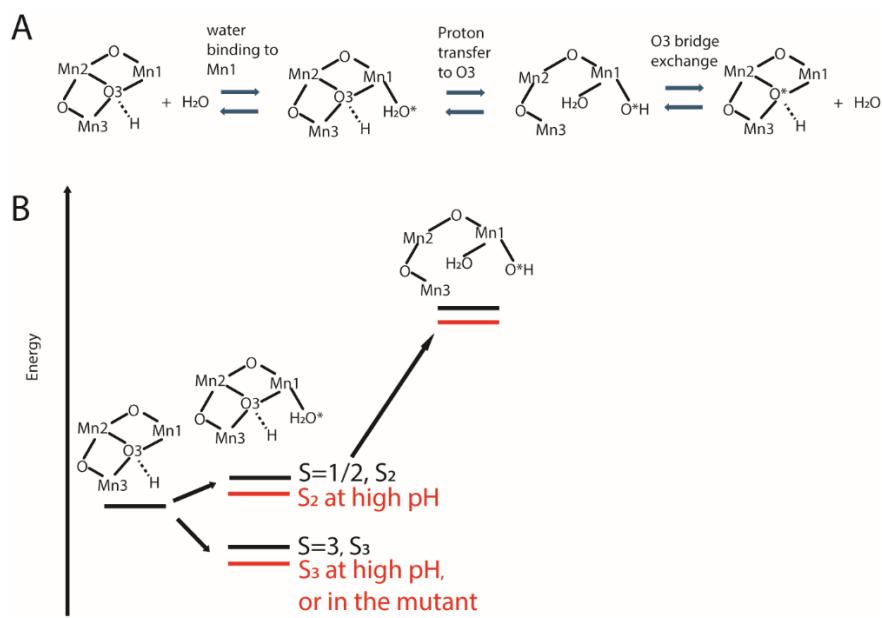

**Figure 2.** A: Mechanism of substrate exchange in the OEC. O3 is assigned as a slow exchanging substrate oxygen, Ws. Steps needed to exchange this oxygen are shown. Note that 6-th ligand on Mn1 is present as a ground state in the S$_3$ -state of the Kok cycle. B: Energy diagram shows how water ligand on Mn1 is over stabilized in the S$_3$ state affecting the population of this state. Changes in the pH and mutations in amino acids in the vicinity of the OEC can affect the substrate exchange rates by changing the energy level of states involved in substrate exchange. The energy difference associated with proton transfer to the O3 bridge (H$_2$O state, large black arrow to the left) for two models of the S$_3$ state with protonated His337 was found to be ~14 kcal/mole, see SI for energy values and coordinates of these models.

Let us see if assignment of the O3 as Ws can satisfy, at least qualitatively, all available observations: i) Ws exchanges the fastest in the $S_0$ state. This state has high content of $Mn^{III}$, thus highest exchange detected in this state is not surprising. ii) Increasing the content of $Mn^{IV}$ in the $S_0$ to $S_1$ oxidation step slows down the exchange while binding of the water to Mn1 is inefficient in the $S_1$ state. Thus, $S_1$ state shows the slowest rate of the O3 exchange. iii) With the transition into $S_2$ state the content of the $Mn^{IV}$ further increases but so is the propensity of this state to bind $H_2O$ at Mn1. Higher population of Mn1-$H_2O$ state allows for a counterintuitive effect of faster substrate exchange despite the increased Mn oxidation state. This was already shown in the earlier calculations where O5 exchange was modeled.[17] iv) $Ca^{2+}$ for $Sr^{2+}$ exchange opens OEC due to coordination of the larger $Sr^{2+}$ ion which supports longer Sr-O distances and increases the population of the Mn1-$H_2O$ state, thus resulting in faster O3 exchange rates.

The overall process can be envisioned as a two-state equilibrium where O3-Mn1-H2O* can either keep exchanging Mn1-H2O* with other waters (fast exchange) or exchange it with O3 (slow exchange).

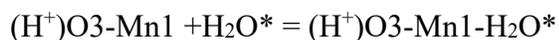

($H^+$)O3-Mn1 +$H_2$O* = ($H^+$)O3-Mn1-$H_2$O*

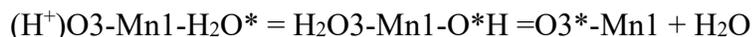

($H^+$)O3-Mn1-$H_2$O* = $H_2$O3-Mn1-O*H =O3*-Mn1 + $H_2$O

\* denotes incoming water

Such exchange mechanism involving O3 as a slow exchange water requires fewer steps than the one proposed earlier for O5 and it also does not require the formation of "closed" cubane state which so far has not been observed experimentally. Below we analyze if available mutagenesis studies are consistent with O3 assessment as critical oxygen center in the OEC.

*Data from available mutagenesis studies do not contradict assignment of the O3 as Ws*

Mutants of His337 were reported: H337R, H337Q, H337N, H337E, H337D, H337Y, H337V, and H337L including the H337F mutant originated by reversion.[25] Only the H337R (Arginine), H337F (phenylalanine), H337Q (glutamine, Gln), and H337N(Asparagine) mutants could be propagated in the absence of glucose. These have shown diminished rates of $O_2$ evolution from ~50-20% of the wild type.[25] Both Arg and Gln are good hydrogen bond donors that can potentially substitute for His337, while Phe was proposed to cause the re-structuring of the OEC coordination with a different amino-acid fragment or water taking the role of the H-bond donor.[25] Thus H-bond to O3 seems to be essential for OEC oxygen evolving function based on the H337 mutagenesis studies.

The D1-V185 residue has been the focus of several recent experimental and theoretical studies as its mutations change the rate of $O_2$ evolution and substrate exchange in an intricate way.[26-29] The V185 residue is located adjacent to the cluster and may also regulate the access to the open coordination site at Mn1. In PSII from *Synechocystis* sp. *PCC 6803,* mutation of the D1-V185 to the polar residue asparagine slows down the release of $O_2$ from 1.2 ms to about 27 ms halftime. On the other hand, mutation of V185 to threonine resulted in only a modest effect on the oxygen release (from 1.2 ms to 1.5 ms).[27] One theoretical study found that by rotating the V185 residue away from the position determined by crystallography the insertion of water from calcium onto Mn1 becomes easier and occurs in the sub-microsecond time-scale rather than the millisecond timescale.[19] Substrate exchange data for D1-V185N mutant strain of *Synechocystis* sp. *PCC 6803* have shown that Ws exchange was faster by a factor ~4.4 for $S_1$ state and ~2 for $S_2$ state.[26]

This was explained by possible stabilization of the Mn1-H$_2$O(or -OH) state by close by asparagine. However, in the S$_3$ state exchange of both slow and fast exchanging substrate oxygens was twice slow. This apparent contradiction can be rationalized by overstabilization of the Mn1-H$_2$O(or -OH) state in the S$_3$ which makes both exchanges with water and O3 slower, **Figure 2B**. Overstabilization of Mn1-OH should also make O-O bond formation harder in agreement with slower O$_2$ evolution rate in this mutant. So, results of substrate exchange in D1-V185N mutant do not contradict the substrate exchange mechanism proposed in **Figure 2A**.

pH dependence of substrate exchange was also analyzed with two pH points of pH=6.0 and pH=8.6 been compared in Ca-PSII and Sr-PSII core preparations from T. elongatus.[16] In the S$_2$ state increasing pH slows fast exchanging substrate slightly and speeds up the exchange of Ws. This is consistent with stabilization of the Mn1-H$_2$O(or -OH) state by high pH, **Figure 2B**. This was noted to be consistent with conversion of the S$_2$ state to high spin (HS)-S$_2$[16] and consistent with original proposal of the HS-S$_2$ as a Mn1-H$_2$O(or -OH) state.[20] In the S$_3$ state, similar to the case of D1-V185N mutant discussed above, the high pH decreases both Wf and Ws exchange rates. Overstabilization of the Mn1-H$_2$O(or -OH) state in the S$_3$ by high pH might be responsible for slower exchange of both O6 and O3 substrate oxygens with outside (unbound) water, **Figure 2B**. High pH=8.6 data might also indicate that electron back donation to Tyrz might not be essential for a substrate exchange in the S$_3$ state. S$_2$Tyrz$^{\bullet+}$ state is more stabilized at basic pH as was confirmed by EPR data[30], while substrate exchange is slowed in the S$_3$ state by basic pH.

*Energetics of the substrate exchange at O3.*

In general, in order to exchange water bound to a Mn center, it should be bound as a terminal and fully protonated ligand on Mn(II) or Mn(III), as Mn(IV) is generally considered to be exchange inert. Additionally, one has to consider that water exchange can occur via two different mechanisms, namely through an associative mechanism, where a new water binds first before the original water ligand leaves, or a dissociative mechanism, where a water is released first before another water binds. As O3 is a ligand to 5-coordinate Mn1 the associative mechanism is a straight forward one. In fact, all previously analyzed schemes of O5 exchange started with water binding to Mn1.[17, 31] We computed several additional states for S$_3$ level of the OEC oxidation and spin state S=3 most consistent with experiments[32], **Figure 1B**, **Table 1**, for more S$_3$ state models presented by our group see [21, 33]. Calculations were done for models with protonated His337 featuring water ligand bound to Mn1 versus the Mn1-OH state with a proton transferred to O3, see **Figure 2A** for the overall exchange proposal. The energy difference associated with proton transfer to the O3 bridge, forming H$_2$O ligand in this position, was found to be ~14 kcal/mole in agreement with characteristics of slow water exchange kinetics. As it was shown earlier that the O5 oxo-bridge protonation step is the highest energy state in the model of the substrate water exchange computed previously at ~18-20 kcal/mol[17], model of the O3 exchange represents a viable option.

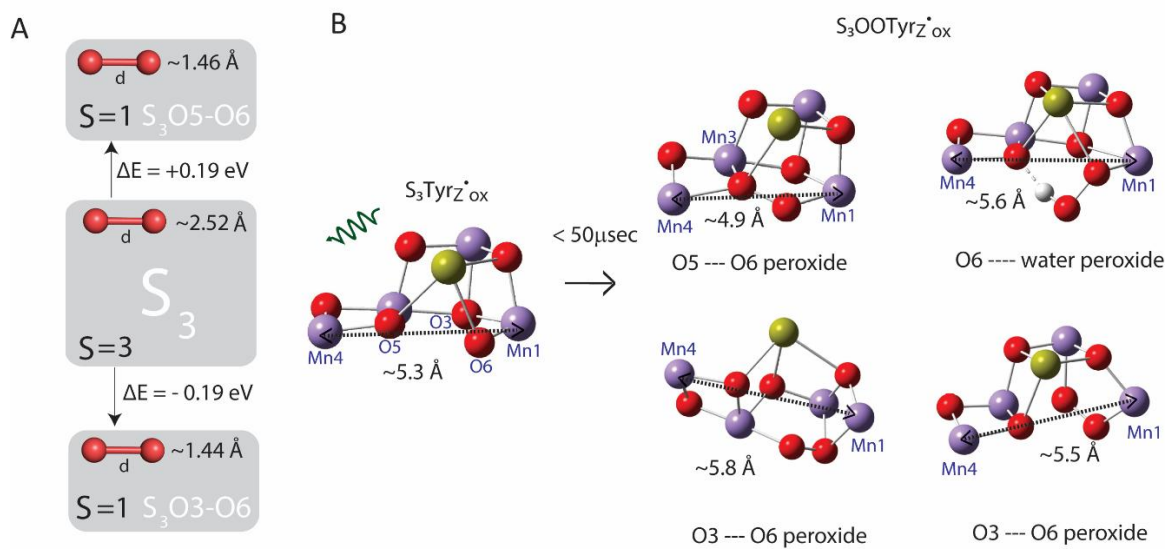

**Figure 3**. A) Scheme comparing the energies of the $S_3$-state model with energies of the corresponding peroxides computed using BP86. O3-O6 peroxide formation lowers the energy when O3 is not H-bonded to His337. B) shows $S_3$-state DFT model featuring the Mn1=O fragment at early time in the $S_3$ to $S_0$ transition after ~ns photoinduced dynamic results in Tyrz oxidation. Right side shows different DFT optimized peroxide models and indicates Mn1-Mn4 distances. All ligands to Mn and Ca except bridging oxygens are removed for clarity. Top line peroxides are expected to be formed at higher energy while the lower ones (O3-O6) are formed with negative ΔE. O3-O6 peroxides satisfy TR-XRD structural trend[4] on elongated Mn1-Mn4 distance in the $S_3$ to $S_0$ transition.

*Mechanism of O-O bond formation with O3 as a substrate oxygen.*

Formation of the O-O bond between the O5 and O6 (or Ox) has been extensively studied by multiple research groups, **Figure 3**.[11, 34-37] It was also used by our group to explain the time-resolved X-ray emission data for the $S_3$ to $S_0$ transition. [21, 38-40] Some PSII TR-XRD data placed these two oxygens in the 1.4-2.2 Å range indicating very close contact even in the $S_3$ state.[5, 41, 42] Thus, this mechanism of the O-O bond formation seemed to be reasonable taking aside the contradictions of the O5 not meeting the properties of the slow exchanging substrate water discussed above. New TR-XRD data[4] also tainted the proposal of the O5-O6 for $O_2$ formation by detecting extension in the Mn1-Mn4 distance of the OEC cluster at all time points in the $S_3$ to $S_0$ transition instead of predicted contraction[40] if O5--O6 peroxide would be an intermediate, **Figure 3**. This contradiction of the O5-O6 peroxide structure and TR-XRD of the $S_3$ to $S_0$ transition was noted and a different, protonated form of peroxo species formed between the O6 and incoming water via water nucleophilic attack mechanism was presented [40], **Figure 3B**. However, no energetic analysis was provided and such state, while in general plausible, does not meet the current understanding of the substrate exchange in the PSII.

O3 similarly to O5 is located in very close proximity of the O6 with O5---O6 vs O3---O6 distances differing at ~0.1-0.3 Å depending on the OEC DFT model of the $S_3$ state. While O5---O6 coupling into the peroxide takes place on the outside the open cubane, O3---O6 coupling will take place almost inside the open cubane unit, assisted by all Mn1-Mn3 ions and $Ca^{2+}$ ion. Several geometries of O3-O6 peroxides models were computed (**Figure 3, Table 1**) and some are found lower in energy in comparison to the $S_3$-state model highlighting a smaller energy gap than reported by us before for O5-O6 peroxides[21].

O3-O6 peroxides also fit better the most recent TR-XRD results showing the elongation of the Mn1-Mn4 distance in the early time points ~50-250 μsec[4], while O5-O6 peroxide should result in shortening of this distance. One more mystery of the TR-XRD results is that second partner oxygen is not visible anywhere in the data but O6 electron density starts decreasing as early as ~250 μsec (when compared to $S_3$ state) and is clearly decreased at 500 μsec.[4] This can be the case if in reality 2$^{nd}$ water (O6,2) also binds to Mn1 and moves pushing the original O6,1 exactly at the time of the O-O bond formation towards the $O_2$ out and taking O3 position after $O_2$ is formed. As this process would lead to minimal displacement O6,1 and O6,2 oxygens would not be distinguishable in TR-XRD, masking 2$^{nd}$ substrate oxygen density dynamics in TR-XRD.

*Analogy of the His337 and the distal His in myoglobin and peroxidases*

Assignment of the O3 as a slow Ws makes it the site of the oxygen production in the PS II. Intriguingly the myoglobin, the other protein which handle $O_2$ molecule, has His site for H-bonding with $O_2$. Thus, when $O_2$ is produced in the PS II, His337 can regulate its reactivity and assist it's exit via H-bond. The distal histidine (His-E7) in myoglobin plays crucial roles in stabilizing bound oxygen, acting as a gate for it's binding, and preventing harmful interactions. The distal E7 histidine in vertebrate myoglobins and hemoglobin has been strongly conserved during evolution and is thought to be important in fine-tuning the $O_2$ affinities of these proteins. A hydrogen bond between the Nε proton of the distal histidine and the second oxygen atom may stabilize $O_2$ bound to the heme iron.[43] The distal histidine (His) which is not unique to myoglobin, is conserved in nearly all globins and in all heme peroxidases, appears to have nearly opposite structural roles in these classes of proteins. In peroxidases, the distal His has the nearly opposite role of activating bound $H_2O_2$ for heterolytic bond cleavage. Prevention of autoxidation would mean prevention of proton shuttling, since protonation of bound $O_2$ is the first step in autoxidation of globins. Note that all His337 mutants have shown lower stability under illumination [25] hinting at potential lack of stabilization mechanism for safe handling of produced $O_2$.

**Summary**

In the OEC structure the bridging O3 is a frustrated center placed in between four dissimilar positive charges: three Mn ions and one proton. Its state is highly modulated from been protonated (estimated pKa~6); been H-bonded with His337 and lacking stabilization of the H-bond, **Figure 1B**. Environment of the His337 Nδ nitrogen is involved with extended H-bond network (**Figure 1C**) and thus can be modulated in the Kok cycle to affect, via Nδ - Nε tautomeric states, the charge on the O3. We show that when O3 lacks the stabilization of the H-bond to His337, the energy of O3-O6 peroxide formation is negative even in the $S_3$ state. In the PS II OEC the mechanism of O-O bond formation between the O3 and O6 activated centers is currently the one which is most consistent with all available experimental data, including the data on substrate exchange, EPR and TR-XRD. The mechanism highlights how the protein environment can steer the O-O bond formation by charge control on the O3 via H-bond and open coordination of Mn1 via 5-6 coordinate conversion dynamics.

**Table 1.** Energies of selected intermediates demonstrating the rearrangement of protons for modeling of the substrate exchange and rearrangement of oxygens for modeling the O-O bond formation. All energies are in Hartree unless energy units are indicated.

| $S_3$-state model with Mn4-$H_2O$, -OH, S=3, charge 2+ and protonated His337 shows proton transfer from Mn1-$H_2O$ to form Mn1-OH;O3H | | | |
|---|---|---|---|
| Initial, E -8387.0807 | Final, E -8387.0583 | ΔE +0.0224 | ΔE +14 kcal/mol |
| $S_3$-state model with Mn4-$H_2O$, -OH, charge -1 and His337-Nδ-H* configuration shows preferential O3-O6 peroxide formation | | | |
| Initial, E, S=3 -8120.085278 | Final, E, S=1 -8120.08625 | ΔE -0.00098 | ΔE -0.02 eV |
|  | Final, E, S=1 -8120.09226 | -0.00700 | -0.19 eV |

\* Lack of the H-bond between the His337-Nε-H and O3 makes His337 drift away from our unrestrained cluster OEC model. Thus, His337 was removed from the model for this set of calculations.

## References

1. Umena, Y., et al., *Crystal structure of oxygen-evolving photosystem II at a resolution of 1.9A*. Nature, 2011. **473**(7345): p. 55-60.
2. Kok, B., B. Forbush, and M. McGloin, *Cooperation of charges in photosynthetic oxygen evolution. I. A linear four step mechanism*. Photochem. Photobiol., 1970. **11**(6): p. 457-75.
3. Hussein, R., et al., *Cryo-electron microscopy reveals hydrogen positions and water networks in photosystem II*. Science, 2024. **384**(6702): p. 1349-1355.
4. Bhowmick, A., et al., *Structural evidence for intermediates during $O_2$ formation in photosystem II*. Nature, 2023. **617**(7961).
5. Li, H.J., et al., *Oxygen-evolving photosystem II structures during $S_1$-$S_2$-$S_3$ transitions*. Nature, 2024. **626**(7999): p. 670-+.
6. Suga, M., et al., *Native structure of photosystem II at 1.95 angstrom resolution viewed by femtosecond X-ray pulses*. Nature, 2015. **517**(7532): p. 99-U265.
7. de Lichtenberg, C., et al., *Assignment of the slowly exchanging substrate water of nature's water-splitting cofactor*. Proceedings of the National Academy of Sciences, 2024. **121**(11): p. e2319374121.
8. Chrysina, M., et al., *Five-coordinate Mn-IV intermediate in the activation of nature's water splitting cofactor*. Proceedings of the National Academy of Sciences of the United States of America, 2019. **116**(34): p. 16841-16846.
9. Pantazis, D.A., *The S-3 State of the Oxygen-Evolving Complex: Overview of Spectroscopy and XFEL Crystallography with a Critical Evaluation of Early-Onset Models for O-O Bond Formation*. Inorganics, 2019. **7**(4): p. 30.
10. Pantazis, D.A., *Missing Pieces in the Puzzle of Biological Water Oxidation*. Acs Catalysis, 2018. **8**(10): p. 9477-9507.

11. Song, Y.T., X.C. Li, and P.E.M. Siegbahn, *Is There a Different Mechanism for Water Oxidation in Higher Plants?* Journal of Physical Chemistry B, 2023. **127**(30): p. 6643-6647.
12. Hendry, G. and T. Wydrzynski, *18O Isotope Exchange Measurements Reveal that Calcium is Involved in the Binding of one Substrate-water Molecule to the Oxygen-Evolving Complex in Photosystem II.* Biochemistry, 2003. **42**(20): p. 6209-6217.
13. Messinger, J., M. Badger, and T. Wydrzynski, *Detection of one slowly exchanging substrate water molecule in the $S_3$ state of Photosystem II.* Proceedings of the National Academy of Sciences, USA, 1995. **92**: p. 3209-3213.
14. Hillier, W. and T. Wydrzynski, *Oxygen ligand exchange at metal sites - implications for the O2 evolving mechanism of photosystem II.* Biochim. Biophys. Acta, 2001. **1503**(1-2): p. 197-209.
15. Nilsson, H., et al., *Substrate-water exchange in photosystem II is arrested before dioxygen formation.* Nature Communications, 2014. **5**: p. 7.
16. de Lichtenberg, C. and J. Messinger, *Substrate water exchange in the S2 state of photosystem II is dependent on the conformation of the Mn4Ca cluster.* Physical Chemistry Chemical Physics, 2020. **22**(23): p. 12894-12908.
17. Siegbahn, P.E.M., *Substrate Water Exchange for the Oxygen Evolving Complex in PSII in the S1, S2, and S3 States.* Journal of the American Chemical Society, 2013. **135**(25): p. 9442-9449.
18. Guo, Y., et al., *The open-cubane oxo-oxyl coupling mechanism dominates photosynthetic oxygen evolution: a comprehensive DFT investigation on O-O bond formation in the S-4 state.* Physical Chemistry Chemical Physics, 2017. **19**(21): p. 13909-13923.
19. Siegbahn, P.E.M., *The S-2 to S-3 transition for water oxidation in PSII (photosystem II), revisited.* Physical Chemistry Chemical Physics, 2018. **20**(35): p. 22926-22931.
20. Pushkar, Y., et al., *Early Binding of Substrate Oxygen Is Responsible for a Spectroscopically Distinct S2 State in Photosystem II.* The Journal of Physical Chemistry Letters, 2019. **10**(17): p. 5284-5291.
21. Pushkar, Y., K.M. Davis, and M. Palenik, *Model of the Oxygen Evolving Complex Which is Highly Predisposed to O–O Bond Formation.* Journal of Physical Chemistry Letters, 2018. **9**: p. 3524-3531.
22. Petrie, S., et al., *The interaction of His337 with the Mn4Ca cluster of photosystem II.* Physical Chemistry Chemical Physics, 2012. **14**(13): p. 4651-4657.
23. Corry, T.A. and P.J. O'Malley, *Proton Isomers Rationalize the High- and Low-Spin Forms of the S2 State Intermediate in the Water-Oxidizing Reaction of Photosystem II.* The Journal of Physical Chemistry Letters, 2019. **10**(17): p. 5226-5230.
24. Mermigki, M.A., M. Drosou, and D.A. Pantazis, *On the nature of high-spin forms in the S2 state of the oxygen-evolving complex.* Chemical Science, 2025. **16**(9): p. 4023-4047.
25. Chu, H.A., A.P. Nguyen, and R.J. Debus, *Amino Acid Residues That Influence the Binding of Manganese or Calcium to Photosystem II .2. the Carboxy-Terminal Domain of the D1 Polypeptide.* Biochemistry, 1995. **34**(17): p. 5859-5882.

26. de Lichtenberg, C., et al., *The D1-V185N mutation alters substrate water exchange by stabilizing alternative structures of the Mn4Ca-cluster in photosystem II*. Biochimica et Biophysica Acta (BBA) - Bioenergetics, 2021. **1862**(1): p. 148319.
27. Dilbeck, P.L., et al., *Perturbing the Water Cavity Surrounding the Manganese Cluster by Mutating the Residue D1-Valine 185 Has a Strong Effect on the Water Oxidation Mechanism of Photosystem II*. Biochemistry, 2013. **52**(39): p. 6824-6833.
28. Bao, H. and R.L. Burnap, *Structural rearrangements preceding dioxygen formation by the water oxidation complex of photosystem II*. Proceedings of the National Academy of Sciences of the United States of America, 2015. **112**(45): p. E6139-E6147.
29. Sugiura, M., et al., *Probing the role of Valine 185 of the D1 protein in the Photosystem II oxygen evolution*. Biochimica et Biophysica Acta (BBA) - Bioenergetics, 2018. **1859**(12): p. 1259-1273.
30. Geijer, P., F. Morvaridi, and S. Styring, *The S3 State of the Oxygen-Evolving Complex in Photosystem II Is Converted to the S2YZ• State at Alkaline pH*. Biochemistry, 2001. **40**(36): p. 10881-10891.
31. Guo, Y., et al., *Theoretical reflections on the structural polymorphism of the oxygen-evolving complex in the S2 state and the correlations to substrate water exchange and water oxidation mechanism in photosynthesis*. Biochimica et Biophysica Acta (BBA) - Bioenergetics, 2017. **1858**(10): p. 833-846.
32. Cox, N., et al., *Electronic structure of the oxygenevolving complex in photosystem II prior to O-O bond formation*. Science, 2014. **345**(6198): p. 804-808.
33. Bury, G. and Y. Pushkar, *Insights from Ca2+→Sr2+ substitution on the mechanism of O-O bond formation in photosystem II*. Photosynthesis Research, 2024. **162**(2): p. 331-351.
34. Siegbahn, P.E.M., *Water oxidation mechanism in photosystem II, including oxidations, proton release pathways, O-O bond formation and O-2 release*. Biochimica Et Biophysica Acta-Bioenergetics, 2013. **1827**(8-9): p. 1003-1019.
35. Rummel, F. and P.J. O'Malley, *How Nature Makes O2: an Electronic Level Mechanism for Water Oxidation in Photosynthesis*. The Journal of Physical Chemistry B, 2022. **126**(41): p. 8214-8221.
36. Yamaguchi, K., et al., *Theory of chemical bonds in metalloenzymes XXI. Possible mechanisms of water oxidation in oxygen evolving complex of photosystem II*. Molecular Physics, 2018. **116**(5-6): p. 717-745.
37. Isobe, H., et al., *Chemical Equilibrium Models for the S-3 State of the Oxygen-Evolving Complex of Photosystem II*. Inorganic Chemistry, 2016. **55**(2): p. 502-511.
38. Davis, K.M., et al., *Rapid Evolution of the Photosystem II Electronic Structure during Water Splitting*. Physical Review X, 2018. **8**(4): p. 041014.
39. Katherine M. Davis, et al., *Rapid Evolution of the Photosystem II Electronic Structure During Water Splitting*. . arXiv:1506.08862, 2015.
40. Jensen, S., et al., *Challenges of observing O-O bond formation in the Mn4Ca cluster of photosystem II*. Chem, 2025.
41. Suga, M., et al., *Light-induced structural changes and the site of O=O bond formation in PSII caught by XFEL*. Nature, 2017. **543**(7643): p. 131-135.

42. Ibrahim, M., et al., *Clear evidence of binding of Ox to the oxygen-evolving complex of photosystem II is best observed in the omit map.* Proceedings of the National Academy of Sciences of the United States of America, 2021. **118**(24).
43. Olson, J.S., et al., *The role of the distal histidine in myoglobin and haemoglobin.* Nature, 1988. **336**(6196): p. 265-266.

Supporting Information for

**New water oxidation mechanism in Photosystem II resolves major experimental controversies.**

Yulia Pushkar

[1] Department of Physics and Astronomy, Purdue University, West Lafayette, Indiana 47907, USA

*Correspondence to:  ypushkar@purdue.edu

**Materials and methods**

For analysis of energies and geometries of PS II intermediates, DFT calculations of molecular models were done with Gaussian09 unrestricted BP86 Becke's 1988 functional[1] with the gradient corrections of Perdew[2] and the def2tzvp basis set[3] used for all atoms, **Tables 1, S1, S2.** The previously discussed models of the low spin (LS) $S_2$ state (S=1/2) and $S_3$-state (S=3) were used for some comparisons.[4-6] The size of our model is maximized to the level when large basis set calculations can be carried out over a practical time period. In the past we documented smaller models results in Davis et al.[7] to be very consistent with later enlarged models[8-10] yielding essentially the same spin configurations, Mn-Mn bond geometries and energetics. At the level of O-O bond formation / spin configuration analysis, much smaller models have proven to be very insightful, see earlier works of Siegbahn[11].

**Table S1.** Representative $^{17}O$ hfs constants in MHz from DFT calculations with UBP86 potential and def2tzvp basis set for all atoms of the **$S_2$ state models (S=1/2)**.

| Oxygen center numbering | $^{17}O$ $a_{iso}$ | $A_{xx}$ | $A_{yy}$ | $A_{zz}$ |
|---|---|---|---|---|
| 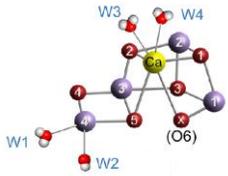 | | | | |
| $S_2$ state (Mn4-OH,-OH, His337 unprotonated) | | | | |
| O5 | -5.5 | 54.4 | -21.1 | -33.2 |
| O4 | 12.7 | 66.4 | -14.2 | -52.2 |
| O2 | -6.1 | 28.3 | -8.3 | -20.0 |
| O1 | -15.3 | 32.7 | 11.3 | -44.0 |
| O3 | -7.9 | 21.1 | 1.0 | -22.2 |
| Mn4-OH | 17.3 | 34.2 | -9.5 | -24.8 |
| Mn4-OH | 20.6 | 23.9 | -8.1 | -15.8 |
| $S_2$ state (Mn4-OH,-H$_2$O, His337 unprotonated) | | | | |
| Mn4-H$_2$O | -4.9 | 1.1 | 0.3 | -1.46 |
| $S_2$-state [Mn4-OH,-OH, His337H$^+$] | | | | |
| O5 | -3.5 | 16.0 | 13.1 | -29.1 |
| O4 | 10.0 | 20.6 | 17.7 | -40.3 |
| O2 | 16.2 | 48.7 | -2.3 | -46.4 |
| O1 | -1.9 | 33.9 | -15.4 | -18.4 |
| O3 | -6.4 | 10.2 | 0.8 | -11.0 |
| Mn4-OH | -23.6 | 33.2 | 12.1 | -45.4 |
| Mn4-OH | -19.8 | 20.8 | 6.3 | -27.1 |

**Table S2.** Mn-Mn and other relevant distances, Å

| | S=1/2, $S_2$ Mn4-OH,-OH, His337 | S=1/2, $S_2$ Mn4-OH,-OH, His337H$^+$ | $S_3$ Mn4-H$_2$O,-OH, Mn1=O, His337 | $S_3$ Mn4-H$_2$O,-OH, Mn1=O, His337H+ |
|---|---|---|---|---|
| Mn4-Mn3 | 2.75 | 2.74 | 2.71 | 2.70 |
| Mn3-Mn2 | 2.78 | 2.87 | 2.78 | 2.87 |
| Mn2-Mn1 | 2.73 | 2.83 | 2.82 | 2.89 |
| Mn1-Mn3 | 3.33 | 3.53 | 3.54 | 3.74 |
| O3-H | 1.88 | 1.05 | 1.87 | 1.03 |
| O3-N$\varepsilon$ | 2.92 | 2.63 | 2.91 | 2.66 |

**Coordinates deposition:**

**S$_2$-state model [Mn4-OH,-OH, His337H$^+$], S=1/2, charge 0.**

| | | | |
|---|---|---|---|
| Mn | -1.15662500 | -3.09941500 | -1.00554800 |
| Mn | -0.50551400 | -0.46816500 | -1.42507600 |
| Mn | -1.24773200 | 2.11990400 | -0.41724000 |
| Mn | 0.18307100 | 0.95834800 | 1.73501200 |
| Ca | -2.94413000 | -0.41400500 | 0.92947000 |
| O | -0.89845200 | -1.46603000 | 0.02910800 |
| O | -1.14737600 | -1.79987200 | -2.35497400 |
| O | -1.95911100 | 0.64010300 | -1.12900800 |
| O | -1.46773000 | 1.54445500 | 1.31437000 |
| O | 0.41785200 | 1.04980200 | -0.23757300 |
| H | -4.66689000 | -4.36980100 | -3.35622700 |
| C | -5.29701600 | -4.51107900 | -2.46599200 |
| H | -6.28150600 | -4.86953800 | -2.80361600 |
| C | -5.44293000 | -3.20980500 | -1.67977400 |
| C | -4.15238000 | -2.58469400 | -1.14406600 |
| O | -4.20552500 | -1.42532100 | -0.68679800 |
| O | -3.11003500 | -3.34376700 | -1.19009000 |
| C | -0.24608400 | 0.62013100 | 6.40302900 |
| C | -1.54438600 | 0.52315700 | 5.60067200 |
| C | -1.38751900 | 0.29822900 | 4.09872000 |
| O | -2.32707200 | -0.20537400 | 3.44779700 |
| O | -0.25145200 | 0.67177500 | 3.59324900 |
| H | 5.93035300 | -2.93656500 | 3.06951500 |
| C | 5.37844300 | -3.38763100 | 2.23295000 |
| C | 5.81236600 | -2.65352600 | 0.96827700 |
| C | 3.84825800 | -3.28647600 | 2.47282400 |
| C | 3.24871400 | -1.91815900 | 2.30419600 |
| N | 3.96569000 | -0.73374000 | 2.42909300 |
| C | 1.94894300 | -1.53341700 | 2.05701400 |
| C | 3.11833400 | 0.30398100 | 2.23520200 |
| N | 1.88680000 | -0.15083900 | 2.02120300 |
| N | 5.61335500 | -3.31373600 | -0.20513400 |
| C | 5.55701600 | -2.61584600 | -1.49321600 |
| H | 6.35333700 | -1.85968000 | -1.48744600 |
| C | 4.18914600 | -1.94704800 | -1.72643200 |
| C | 3.03902600 | -2.95782100 | -1.79545900 |
| C | 1.63725900 | -2.43041100 | -1.48975700 |
| O | 1.38861000 | -1.19780100 | -1.68492800 |
| O | 0.83058800 | -3.31286400 | -1.07425100 |
| C | 5.68766700 | 4.02952300 | -2.01344100 |
| C | 4.50419300 | 3.13100600 | -1.88688000 |
| C | 3.54310200 | 2.99280100 | -0.90898700 |
| C | 3.03040300 | 1.53985900 | -2.43540200 |
| N | 2.64066700 | 2.00695500 | -1.25981200 |
| H | 0.93858900 | 7.35968900 | 1.11761400 |
| C | 0.61011100 | 6.39160400 | 0.71086800 |
| H | -0.47776500 | 6.42565700 | 0.56809500 |
| C | 1.00416600 | 5.25613100 | 1.65366800 |
| C | 0.57018200 | 3.85904600 | 1.21693700 |
| O | -0.26492800 | 3.77853900 | 0.26195900 |
| O | 1.07472000 | 2.87623800 | 1.85907400 |
| C | -4.01160300 | 2.77707700 | -0.01045700 |
| O | -4.25413000 | 1.70759000 | 0.57929700 |
| O | -2.85524900 | 3.13157800 | -0.49446600 |
| C | -1.38466800 | 3.46767900 | -4.95399600 |
| C | -0.19492800 | 2.58756800 | -4.55281300 |
| C | -0.34553200 | 1.95373000 | -3.17388100 |
| O | -0.73614400 | 2.73446400 | -2.23775900 |
| O | -0.04344800 | 0.72751900 | -3.05270300 |
| O | 6.25309300 | -1.49456800 | 1.01647600 |
| H | -4.83943900 | -5.29940200 | -1.85262200 |
| H | 5.20946000 | -4.24444600 | -0.15484300 |
| H | 1.31224400 | 1.46602400 | -0.60508600 |
| H | -5.93260800 | -2.42863400 | -2.28212100 |
| H | -6.09210300 | -3.35400600 | -0.79934600 |
| H | 5.74679700 | 4.68812600 | -1.13762300 |
| H | 6.63180800 | 3.46398200 | -2.07055600 |
| H | 3.44647500 | 3.54554600 | 0.02034400 |
| H | 2.52783000 | 0.75295600 | -2.99082900 |
| H | 3.40873300 | 1.34880400 | 2.24455800 |
| H | 1.06773700 | -2.14470400 | 1.87713100 |
| H | 4.98892900 | -0.68677000 | 2.34800500 |
| H | 5.78586700 | -3.34703100 | -2.28357500 |
| C | -5.11224600 | 3.79382000 | -0.25243700 |
| H | -4.70981000 | 4.81232600 | -0.30810100 |
| H | -5.58797000 | 3.55981500 | -1.21745600 |
| H | -5.87504800 | 3.71682100 | 0.53177900 |
| H | -2.30819300 | 2.87443800 | -5.01783500 |
| H | -1.54523800 | 4.26588600 | -4.21773800 |
| H | -0.02883200 | 1.77763900 | -5.27636200 |
| H | 0.72456300 | 3.19859100 | -4.52262900 |
| H | 0.33640200 | -0.30977000 | 6.33039200 |
| H | -2.19857000 | -0.27584800 | 5.98039300 |
| H | 0.38652100 | 1.43908600 | 6.03638100 |
| H | -2.12337400 | 1.45880600 | 5.70129600 |
| H | 3.64732500 | -3.65857100 | 3.49215400 |
| H | 5.65650700 | -4.45258500 | 2.19560200 |
| H | 3.31077700 | -3.96864200 | 1.79669500 |
| H | 1.06526500 | 6.26310400 | -0.28131400 |
| H | 0.56080000 | 5.40616200 | 2.65366700 |
| H | 2.09235600 | 5.22539700 | 1.82443100 |
| H | 4.01091200 | -1.22447300 | -0.91586600 |
| H | 3.19539500 | -3.79381300 | -1.09779100 |
| H | 2.97785800 | -3.41689800 | -2.79746500 |
| H | 4.23966500 | -1.35968600 | -2.65620700 |
| O | -2.96302400 | -2.63266000 | 1.95547900 |
| H | -2.68944800 | -2.54391700 | 2.88351900 |
| O | -1.10781600 | -4.61675600 | -2.05365400 |
| H | -2.28967200 | -3.25896400 | 1.51035200 |
| O | -1.18227400 | -4.03902000 | 0.65278000 |
| H | -5.12102200 | 0.77869100 | 2.11836000 |
| O | -4.94640200 | -0.10753200 | 2.49931200 |
| H | -4.35309000 | 0.03802100 | 3.26484900 |
| H | -2.01876500 | -4.72383700 | -2.39010600 |
| H | -1.38605900 | -4.96091900 | 0.40682800 |
| H | -0.46525900 | 0.80005800 | 7.46593800 |
| H | 5.62545300 | 4.66975600 | -2.90826100 |
| H | -1.20721900 | 3.92719300 | -5.93744300 |
| N | 4.15377300 | 2.19283800 | -2.84897400 |
| H | 4.64474600 | 2.02829400 | -3.72193900 |

**$S_3$-state model [Mn4-OH2,-OH, Mn1=O, His337H$^+$], S=3, charge 0.**

| | | | |
|---|---|---|---|
| C | -5.99325300 | -2.40231500 | -0.69733000 |
| C | -4.59677100 | -2.20161400 | -0.10613300 |
| O | -4.37629300 | -1.19654400 | 0.59375900 |
| O | -3.72686700 | -3.11935700 | -0.38892400 |
| C | -1.11876300 | 3.77956200 | 4.57049500 |
| C | -1.14133600 | 2.79778100 | 3.39647700 |
| O | -2.14338700 | 2.10597200 | 3.16129000 |
| O | -0.01937400 | 2.81029300 | 2.71880600 |
| C | 5.86314900 | -2.06051200 | 1.47012000 |
| O | 6.28279400 | -1.53198200 | 0.43069200 |
| C | 4.79396800 | -0.95604200 | 3.57594700 |
| C | 3.93926200 | 0.03139400 | 2.84528400 |
| C | 2.59469200 | 0.02335500 | 2.54439500 |
| C | 3.35600100 | 1.93213600 | 1.82177800 |
| N | 2.25431100 | 1.20111200 | 1.91102000 |
| N | 5.20923400 | -3.25325000 | 1.47496200 |
| C | 3.58597400 | -3.69213900 | -0.40275700 |
| C | 2.37102200 | -3.95648100 | 0.48898300 |
| C | 1.06558500 | -3.32424400 | 0.00862900 |
| O | 1.12368200 | -2.21714500 | -0.59795700 |
| O | 0.01168200 | -3.97751900 | 0.32819700 |
| C  |  4.94932500 |  0.88165700 | -2.68406500 |
| C  |  3.64279500 |  1.32367800 | -2.77352700 |
| C  |  3.73018300 |  0.09607300 | -1.00068100 |
| N  |  2.89578500 |  0.82744300 | -1.72248400 |
| C  |  1.54005900 |  5.15099100 | -1.15881700 |
| C  |  0.89306400 |  3.82058400 | -0.77820600 |
| O  |  0.03138200 |  3.34029900 | -1.58753900 |
| O  |  1.27124900 |  3.28705100 |  0.30704000 |
| C  | -3.76340700 |  2.59995000 | -1.13171500 |
| O  | -3.99828600 |  1.94138300 | -0.10465900 |
| O  | -2.64579800 |  2.60290500 | -1.81151400 |
| Ca | -2.54682300 |  0.32669700 |  1.48992100 |
| O  | -1.28432200 |  2.09577500 |  0.44206500 |
| O  | -1.91905800 |  0.07871900 | -1.24786500 |
| O  |  0.48355900 |  0.62277500 | -0.60966400 |
| O  | -1.36374600 | -2.65832800 | -1.61747400 |
| O  | -1.31254100 | -1.65109500 |  0.55991000 |
| O  | -0.24338300 |  0.25966700 |  2.01563000 |
| Mn |  0.31247300 |  1.65319800 |  1.24672100 |
| Mn | -1.06222500 |  1.68163000 | -1.29385300 |
| Mn | -0.58572500 | -1.14194100 | -1.07639000 |
| Mn | -1.80472800 | -3.26851200 |  0.06517000 |
| O  | -2.23537900 | -5.14744700 | -0.99091800 |
| O  | -4.98961300 |  0.83159700 |  2.24785100 |
| O  | -2.29067700 | -3.84973900 |  1.79498300 |
| O  | -2.71804500 | -1.48046600 |  3.19592000 |
| C  |  0.21537200 |  0.04028800 | -5.11999000 |
| C | -0.10931300 | 0.18504300 | -3.63814400 |
| O | -0.54080900 | 1.30273500 | -3.22279200 |
| O | 0.09181800 | -0.85687000 | -2.92552600 |
| H | -6.70111400 | -2.33232400 | 0.14497400 |
| H | -6.18403800 | -1.50887900 | -1.31448200 |
| H | -0.25742400 | 3.50753300 | 5.20303400 |
| H | -0.87754300 | 4.77359000 | 4.15924100 |
| H | 4.15676800 | -1.82141400 | 3.81351900 |
| H | 5.10183000 | -0.53968400 | 4.55165400 |
| H | 1.82230300 | -0.71419600 | 2.74453500 |
| H | 3.41487200 | 2.91307600 | 1.36256900 |
| H | 3.50839700 | -4.28531500 | -1.32864300 |
| H | 3.59067900 | -2.63844500 | -0.71244700 |
| H | 2.19610800 | -5.03099800 | 0.64519600 |
| H | 2.53435100 | -3.51570100 | 1.48945400 |
| H | 4.90896500 | -3.62771500 | 2.36966900 |
| H | 3.20386100 | 1.96721700 | -3.53028100 |
| H | 1.36820000 | 0.76286700 | -1.12976200 |
| H | 3.46948600 | -0.44851200 | -0.09874900 |
| H | 1.44672400 | 5.80567800 | -0.27712400 |
| H | 2.62070500 | 4.95397800 | -1.26580200 |
| H | -0.39782300 | -0.79894600 | -5.48810200 |
| H | 1.25783100 | -0.31316000 | -5.18180100 |
| H | -1.70427400 | -4.92388900 | -1.78374100 |
| H | -3.14648300 | -4.85953800 | -1.22458400 |
| H | -5.22938400 | 0.10461500 | 1.62738700 |
| H | -5.02641300 | 1.62162600 | 1.66959200 |
| H | -3.14883500 | -4.31067000 | 1.76821300 |
| H | -2.53566500 | -2.35859600 | 2.74575300 |
| H | -3.59922300 | -1.56814500 | 3.59547000 |
| C | 6.14401800 | 1.11227700 | -3.54769000 |
| H | 6.96617100 | 1.59539000 | -2.99508400 |
| H | 6.53551100 | 0.17166300 | -3.96784900 |
| H | 5.87555800 | 1.76659500 | -4.38723500 |
| C | -0.00294100 | 1.29561800 | -5.96171400 |
| H | 0.62206900 | 2.12804500 | -5.61035000 |
| H | 0.24657900 | 1.09742300 | -7.01484600 |
| H | -1.04769300 | 1.63072600 | -5.91137700 |
| C | -6.22519300 | -3.67860600 | -1.50367700 |
| H | -6.06671800 | -4.57788800 | -0.89076900 |
| H | -7.25745800 | -3.71061000 | -1.88300000 |
| H | -5.54826600 | -3.73232200 | -2.36805900 |
| C | -2.41266800 | 3.81158800 | 5.38062800 |
| H | -2.64127300 | 2.82256500 | 5.80060000 |
| H | -2.33027800 | 4.53135300 | 6.20922000 |
| H | -3.26604700 | 4.10225200 | 4.75278200 |
| C | -4.81024700 | 3.53750200 | -1.71408100 |
| H | -4.58129600 | 4.56406200 | -1.38868100 |
| H | -4.78215800 | 3.52235100 | -2.81110800 |
| H | -5.80743800 | 3.26540000 | -1.34847200 |
| C | 0.97999300 | 5.82263400 | -2.41108700 |
| H | 1.50486900 | 6.76993100 | -2.60688000 |
| H | 1.08918400 | 5.17724500 | -3.29295700 |
| H | -0.09097500 | 6.03950400 | -2.29834000 |
| C | 4.92648100 | -4.03389200 | 0.26619400 |
| H | 4.96174100 | -5.09775500 | 0.54904600 |
| H | 5.74940700 | -3.84526200 | -0.43761000 |
| C | 6.07918600 | -1.41585800 | 2.84124500 |
| H | 6.75389900 | -0.56540400 | 2.67133200 |
| H | 6.61827300 | -2.12243000 | 3.49433300 |
| N | 4.39820800 | 1.26168200 | 2.38153500 |
| H | 5.35438800 | 1.59941900 | 2.42009000 |
| N | 4.97757400 | 0.10056300 | -1.54011700 |
| H | 5.75871700 | -0.41637300 | -1.12746800 |

**$S_3$-state model [Mn4-$H_2O$, -OH; Mn1-$H_2O$; His337$H^+$], S=3, charge 2+**

| | | | |
|---|---|---|---|
| C | -6.17923200 | 1.99915900 | 1.30659500 |
| C | -4.81902800 | 1.79005200 | 0.65246000 |
| O | -4.62705000 | 0.78166700 | -0.05823600 |
| O | -3.93497800 | 2.70974700 | 0.88658800 |
| C | -1.16123400 | -2.89052700 | -4.97159600 |
| C | -1.10260500 | -2.02641600 | -3.72346100 |
| O | -2.10433000 | -1.44735900 | -3.26328400 |
| O | 0.10527300 | -1.94395300 | -3.20231800 |
| C | 5.39391700 | 2.83463400 | -1.15477200 |
| O | 5.43888600 | 1.95430400 | -0.28066400 |
| C | 4.66454700 | 2.09208500 | -3.50403600 |
| C | 3.88477600 | 0.91264000 | -3.01749000 |
| C | 2.63789600 | 0.83295900 | -2.44317200 |
| C | 3.36914200 | -1.21638800 | -2.57255800 |
| N | 2.32972200 | -0.48792500 | -2.16744300 |
| N  |  4.92183000 |  4.08443800 | -0.90150500 |
| C  |  3.22556800 |  3.92628500 |  0.96142000 |
| C  |  2.00095900 |  4.24817500 |  0.10072000 |
| C  |  0.78011900 |  3.38655100 |  0.38989300 |
| O  |  0.95939000 |  2.19630900 |  0.79732900 |
| O  | -0.35233800 |  3.93031400 |  0.14026100 |
| C  |  4.64770600 | -1.82421600 |  2.85989400 |
| C  |  3.33305700 | -1.92370500 |  2.47184100 |
| C  |  4.20186900 | -0.22458300 |  1.33976500 |
| N  |  3.08710700 | -0.92887100 |  1.53925400 |
| C  |  2.09092500 | -4.97679900 | -0.24178000 |
| C  |  1.32595300 | -3.67092700 | -0.14223900 |
| O  |  0.58880400 | -3.44281400 |  0.85126900 |
| O  |  1.49995700 | -2.84752500 | -1.12495600 |
| C  | -3.23004200 | -3.20242800 |  1.03914800 |
| O  | -3.70846700 | -2.44791400 |  0.17061600 |
| O  | -2.00639000 | -3.12822400 |  1.50167300 |
| Ca | -3.04934900 | -0.58249600 | -1.10015700 |
| O  | -1.04087600 | -1.93502200 | -0.80192200 |
| O  | -1.75171500 | -0.46684800 |  1.33475300 |
| O  |  0.65806600 | -0.57726100 |  0.35206400 |
| O  | -1.49538200 |  2.26166200 |  1.92497200 |
| O  | -1.48166300 |  1.45935200 | -0.32100500 |
| Mn |  0.48740500 | -1.24348000 | -1.46118900 |
| Mn | -0.68122400 | -1.86599300 |  1.02068300 |
| Mn | -0.56178600 |  0.92678700 |  1.22997000 |
| Mn | -2.09987200 |  3.02359200 |  0.35924300 |
| O | -2.61300300 | 4.84150300 | 1.26382800 |
| O | -5.21584000 | -0.91085000 | -2.18533300 |
| O | -2.62700000 | 3.56190800 | -1.36217400 |
| O | -2.86848500 | 1.29553200 | -2.73215800 |
| C | 0.50698700 | -0.72174300 | 4.99977400 |
| C | 0.22035900 | -0.67605100 | 3.51229100 |
| O | -0.00936600 | -1.76843900 | 2.89466700 |
| O | 0.23569300 | 0.47970500 | 2.96223000 |
| H | -6.91008200 | 2.02841400 | 0.48049900 |
| H | -6.39464100 | 1.06561300 | 1.85200100 |
| H | -0.34871500 | -2.55928300 | -5.63849200 |
| H | -0.87519700 | -3.90946700 | -4.65866100 |
| H | 3.95541800 | 2.92232600 | -3.63765400 |
| H | 5.07634300 | 1.88310100 | -4.50493500 |
| H | 1.93395500 | 1.63004200 | -2.24163500 |
| H | 3.43652600 | -2.29728800 | -2.52874800 |
| H | 3.06956300 | 4.29307100 | 1.98830300 |
| H | 3.34504400 | 2.83821000 | 1.03108800 |
| H | 1.69990900 | 5.30324200 | 0.17386000 |
| H | 2.23524200 | 4.07265700 | -0.96559700 |
| H | 5.01030600 | 4.78376200 | -1.63316300 |
| H | 2.55465900 | -2.61101200 | 2.78356000 |
| H | 2.15099500 | -0.74708000 | 1.06054200 |
| H | 4.35876900 | 0.61992000 | 0.66613600 |
| H | 1.83371100 | -5.40937800 | -1.22297300 |
| H | 3.16156100 | -4.71929300 | -0.31401300 |
| H | -0.23431100 | -0.04904200 | 5.46340900 |
| H | 1.47646900 | -0.21901000 | 5.15494700 |
| H | -1.87549500 | 5.23313700 | 1.76948700 |
| H | -3.33101300 | 4.62675300 | 1.89457000 |
| H | -5.81143000 | -0.51891000 | -1.51700700 |
| H | -5.61771000 | -1.75570400 | -2.45149300 |
| H | -3.31349800 | 4.25444300 | -1.35322100 |
| H | -2.86815500 | 2.21976700 | -2.31526300 |
| H | -3.46639000 | 1.31253000 | -3.49825500 |
| C | 5.46215700 | -2.61009700 | 3.82786500 |
| H | 6.32752300 | -3.08217600 | 3.33820100 |
| H | 5.84120500 | -1.97557000 | 4.64349700 |
| H | 4.85218400 | -3.40427200 | 4.27435600 |
| C | 0.47313000 | -2.10633900 | 5.64204500 |
| H | 1.23565600 | -2.77408400 | 5.21787900 |
| H | 0.66151000 | -2.02120100 | 6.72090700 |
| H | -0.50324400 | -2.58813600 | 5.50272200 |
| C | -6.33592800 | 3.21610400 | 2.21493000 |
| H | -6.17316700 | 4.15464900 | 1.66634300 |
| H | -7.35346200 | 3.24855100 | 2.62802600 |
| H | -5.63583900 | 3.17791900 | 3.06154500 |
| C | -2.51188500 | -2.88761000 | -5.68300300 |
| H | -2.78937300 | -1.87780600 | -6.01448600 |
| H | -2.46990400 | -3.53659400 | -6.56841800 |
| H | -3.31022000 | -3.25824200 | -5.02649300 |
| C | -4.04032200 | -4.32943400 | 1.63472100 |
| H | -3.76865800 | -5.26619800 | 1.12357500 |
| H | -3.81240600 | -4.45644500 | 2.70000200 |
| H | -5.10960800 | -4.14476900 | 1.48388800 |
| C | 1.82811600 | -5.96779600 | 0.88997600 |
| H | 2.40735200 | -6.88563600 | 0.72094900 |
| H | 2.11732000 | -5.55646500 | 1.86616500 |
| H | 0.76585800 | -6.23853500 | 0.94466700 |
| C | 4.53473200 | 4.53903100 | 0.44122800 |
| H | 4.46558800 | 5.63536300 | 0.39829500 |
| H | 5.34579800 | 4.28614400 | 1.14143200 |
| C | 5.83336700 | 2.55109800 | -2.58721700 |
| H | 6.60343100 | 1.76963100 | -2.52888000 |
| H | 6.29936200 | 3.43778000 | -3.04435100 |
| N | 4.32353400 | -0.40387700 | -3.08596000 |
| H | 5.20327000 | -0.72031000 | -3.48527500 |
| N | 5.15551300 | -0.75301300 | 2.13000900 |
| H | 6.10896700 | -0.40034000 | 2.17599700 |
| H | -0.69485400 | 1.06195600 | -1.41442300 |
| O | -0.21298700 | 0.48752900 | -2.15948700 |
| H | -0.91581400 | 0.38637100 | -2.84757900 |

**S$_3$-state model [Mn4-H$_2$O,-OH; Mn1-OH; O3H; His337H$^+$], S=3, charge 2+.**

| | | | |
|---|---:|---:|---:|
| C | -6.13367900 | 2.14735400 | 1.27653300 |
| C | -4.77809500 | 1.90729200 | 0.62343700 |
| O | -4.60689500 | 0.89183300 | -0.08248000 |
| O | -3.87534400 | 2.80974900 | 0.85324500 |
| C | -1.54704500 | -3.40856300 | -4.15129000 |
| C | -1.14105000 | -2.00526700 | -3.73428100 |
| O | -1.88454700 | -1.01905400 | -3.89276800 |
| O | 0.06830000 | -1.94514500 | -3.21349400 |
| C | 5.45401700 | 2.73353600 | -1.18838200 |
| O | 5.48096900 | 1.85660900 | -0.31012300 |
| C | 4.70941200 | 1.99501800 | -3.53413500 |
| C | 3.90563500 | 0.83413400 | -3.04204600 |
| C | 2.65741900 | 0.78277000 | -2.46739800 |
| C | 3.34645100 | -1.28173900 | -2.58707500 |
| N | 2.32222400 | -0.53019500 | -2.18544000 |
| N | 5.00769100 | 3.99394700 | -0.94103900 |
| C | 3.30866200 | 3.87943900 | 0.92255800 |
| C | 2.09086000 | 4.22231900 | 0.06030700 |
| C | 0.85261600 | 3.38730300 | 0.35350900 |
| O | 1.00745200 | 2.19557200 | 0.76657100 |
| O | -0.26846100 | 3.95300400 | 0.10127400 |
| C | 4.61262600 | -1.88999900 | 2.84823000 |
| C | 3.29618700 | -1.96432300 | 2.46060900 |
| C | 4.19960800 | -0.28875500 | 1.32054500 |
| N | 3.07064200 | -0.96906900 | 1.52332400 |
| C | 1.99149300 | -5.00407400 | -0.23859400 |
| C | 1.25348400 | -3.68232500 | -0.14524900 |
| O | 0.52123500 | -3.43444200 | 0.84714600 |
| O | 1.44428400 | -2.86731900 | -1.13184100 |
| C | -3.29186200 | -3.11486000 | 1.03376400 |
| O | -3.75476000 | -2.35480000 | 0.16166100 |
| O | -2.06691500 | -3.06359600 | 1.49597200 |
| Ca | -3.05758600 | -0.50932500 | -1.11789200 |
| O | -1.07726900 | -1.90135500 | -0.81320400 |
| O | -1.75769400 | -0.40882400 | 1.31648600 |
| O | 0.64925000 | -0.57330400 | 0.33440700 |
| O | -1.44538900 | 2.31661000 | 1.89381300 |
| O | -1.44828100 | 1.50358300 | -0.34834700 |
| Mn | 0.46483700 | -1.24444100 | -1.47568300 |
| Mn | -0.71615900 | -1.83110900 | 1.00906600 |
| Mn | -0.53942600 | 0.95959000 | 1.20515700 |
| Mn | -2.03421900 | 3.08337600 | 0.32448200 |
| O | -2.50988000 | 4.91568800 | 1.22044900 |
| O | -5.23042800 | -0.79827800 | -2.20157400 |
| O | -2.55030100 | 3.62425600 | -1.39947700 |
| O | -2.83832900 | 1.35686400 | -2.75874400 |
| C | 0.49553900 | -0.69269300 | 4.98274300 |
| C | 0.20981300 | -0.64815800 | 3.49505200 |
| O | -0.04231900 | -1.73850800 | 2.88258900 |
| O | 0.24882400 | 0.50442800 | 2.93953600 |
| H | -6.86382900 | 2.18769600 | 0.45028400 |
| H | -6.36816400 | 1.22101200 | 1.82633700 |
| H | -0.72130600 | -3.82443700 | -4.75117600 |
| H | -1.56281900 | -4.01509700 | -3.22941600 |
| H | 4.01746000 | 2.83899400 | -3.67169700 |
| H | 5.11676800 | 1.77290100 | -4.53402200 |
| H | 1.96999600 | 1.59507300 | -2.26965200 |
| H | 3.39164400 | -2.36357500 | -2.53815500 |
| H | 3.16028300 | 4.25419700 | 1.94769100 |
| H | 3.40579100 | 2.78948200 | 0.99737000 |
| H | 1.81152800 | 5.28367500 | 0.12845000 |
| H | 2.32142300 | 4.03699400 | -1.00516100 |
| H | 5.11045100 | 4.68784300 | -1.67599100 |
| H | 2.50387000 | -2.63403200 | 2.77554800 |
| H | 2.13842600 | -0.77037200 | 1.04372800 |
| H | 4.37376000 | 0.54915800 | 0.64293800 |
| H | 1.72539300 | -5.43591600 | -1.21774000 |
| H | 3.06718200 | -4.76893600 | -0.31200800 |
| H | -0.23176900 | -0.00273900 | 5.44317000 |
| H | 1.47514300 | -0.20923100 | 5.13557000 |
| H | -1.76445900 | 5.29449100 | 1.72427600 |
| H | -3.23210500 | 4.71869900 | 1.85217600 |
| H | -5.81780700 | -0.39104600 | -1.53512700 |
| H | -5.64956700 | -1.63595600 | -2.46375200 |
| H | -3.22244300 | 4.33076600 | -1.39381900 |
| H | -2.81900700 | 2.28285700 | -2.34622100 |
| H | -3.43580900 | 1.38250700 | -3.52493200 |
| C | 5.41084200 | -2.68784400 | 3.81993000 |
| H | 6.26630700 | -3.17988900 | 3.33253100 |
| H | 5.80288300 | -2.05738100 | 4.63256700 |
| H | 4.78473000 | -3.46721700 | 4.27015000 |
| C | 0.43332000 | -2.07325200 | 5.63154900 |
| H | 1.18195600 | -2.75850000 | 5.21056800 |
| H | 0.62347700 | -1.98690100 | 6.71000300 |
| H | -0.55274400 | -2.53556600 | 5.49447300 |
| C | -6.26531200 | 3.37153700 | 2.17910200 |
| H | -6.08336300 | 4.30394100 | 1.62609100 |
| H | -7.28193900 | 3.42680800 | 2.59200700 |
| H | -5.56609900 | 3.32299500 | 3.02591000 |
| C | -2.87948900 | -3.49041400 | -4.89173400 |
| H | -2.85967200 | -2.90468500 | -5.82077300 |
| H | -3.09902800 | -4.53471100 | -5.15278800 |
| H | -3.70444800 | -3.11328900 | -4.27291900 |
| C | -4.12505800 | -4.22217500 | 1.63463000 |
| H | -3.87270700 | -5.16672100 | 1.12792500 |
| H | -3.89972800 | -4.34880100 | 2.70050600 |
| H | -5.19034000 | -4.01632200 | 1.48289100 |
| C | 1.70847700 | -5.98411100 | 0.89782400 |
| H | 2.26874700 | -6.91443200 | 0.73315500 |
| H | 2.00612400 | -5.57419300 | 1.87206800 |
| H | 0.64089100 | -6.23273500 | 0.95376000 |
| C | 4.63009000 | 4.46272700 | 0.39951800 |
| H | 4.58345400 | 5.56003200 | 0.35140300 |
| H | 5.43584100 | 4.19656200 | 1.10093500 |
| C | 5.88746400 | 2.43427700 | -2.61945700 |
| H | 6.64133400 | 1.63745800 | -2.55740300 |
| H | 6.37152500 | 3.30904100 | -3.08076100 |
| N | 4.31728200 | -0.49141800 | -3.10428000 |
| H | 5.19031300 | -0.82771900 | -3.50206600 |
| N | 5.14225900 | -0.83290200 | 2.11330800 |
| H | 6.10275200 | -0.49965400 | 2.15765900 |
| H | 1.05674500 | 0.36101500 | 0.18643300 |
| O | -0.16871200 | 0.41545800 | -2.14899500 |
| H | -0.90473200 | 0.40729100 | -2.86980400 |

**S$_3$-state model with Mn4-H$_2$O, -OH, charge -1. S=3 and His337-Nδ-H* configuration**

\* Lack of the H-bond between the His337-Nε-H and O3 makes His337 drift away from our unrestrained cluster OEC model. Thus, His337 was removed from the model for this set of calculations.

| | | | |
|---|---|---|---|
| C | 5.11202800 | -3.30343000 | -0.50451700 |
| C | 3.68874500 | -2.83088400 | -0.82060300 |
| O | 3.54594500 | -1.77003700 | -1.46397100 |
| O | 2.73339100 | -3.57786000 | -0.38555500 |
| C | 0.53620400 | 4.26108300 | -3.94863200 |
| C | 0.55760500 | 3.05386600 | -2.99764200 |
| O | 1.43193000 | 2.17191800 | -3.14584200 |
| O | -0.37338900 | 3.10075300 | -2.09968100 |
| C | -6.77149700 | -0.71428400 | -0.20978100 |
| O | -6.93611000 | -0.07588300 | 0.83821400 |
| C | -5.71394900 | 0.20288900 | -2.37704200 |
| C | -4.59034700 | 0.82439300 | -1.59800000 |
| C | -3.23601400 | 0.84204500 | -1.84516000 |
| C | -3.52670000 | 1.99391100 | -0.02659800 |
| N | -2.59384700 | 1.57108400 | -0.86719900 |
| N | -6.35442700 | -2.01150200 | -0.23163700 |
| C | -4.36280300 | -2.52842800 | 1.20863700 |
| C | -3.47770000 | -3.12340100 | 0.11156100 |
| C | -1.98939900 | -2.79791400 | 0.26690000 |
| O | -1.67236100 | -1.79703900 | 0.95624200 |
| O | -1.20262500 | -3.59100800 | -0.36745300 |
| C | -0.60416300 | 5.23147600 | 1.98504600 |
| C | -0.27761800 | 3.82812700 | 1.46305600 |
| O | 0.71523100 | 3.23226800 | 1.99101000 |
| O  | -1.02200200 | 3.38029200 | 0.54251800 |
| C  | 4.09792600 | 1.86167000 | 0.68016000 |
| O  | 3.97435300 | 1.25348400 | -0.40820800 |
| O  | 3.20119700 | 1.99487800 | 1.59862600 |
| Ca | 1.95404700 | 0.20649400 | -1.78969500 |
| O  | 1.27533300 | 1.92811500 | -0.31370600 |
| O  | 1.87655100 | -0.32730800 | 1.23317200 |
| O  | -0.38340400 | 0.70299800 | 0.97812400 |
| O  | 0.77414900 | -2.90631100 | 1.36588100 |
| O  | 0.53881900 | -1.56149000 | -0.65688700 |
| O  | -0.40299000 | 0.47932900 | -1.79263900 |
| Mn | -0.51344600 | 1.71202300 | -0.69872400 |
| Mn | 1.33655200 | 1.39269700 | 1.40223300 |
| Mn | 0.30177200 | -1.13221600 | 1.12484000 |
| Mn | 0.74083800 | -3.29296300 | -0.43197900 |
| O  | 0.97273600 | -5.35520600 | 0.48791100 |
| O  | 4.35632400 | 0.33699400 | -2.93519700 |
| O  | 0.77391000 | -3.76528200 | -2.28674100 |
| O  | 1.41717000 | -1.44288800 | -3.60330300 |
| C  | 0.31025600 | -0.23321500 | 5.33222500 |
| C  | 0.46548000 | -0.04148200 | 3.82173900 |
| O  | 1.07114400 | 0.99573500 | 3.41732000 |
| O  | -0.03298600 | -0.96872900 | 3.10294000 |
| H  | 5.66627900 | -3.30294600 | -1.45760200 |
| H  | 5.56219800 | -2.49762900 | 0.09861600 |
| H  | -0.51419000 | 4.46918700 | -4.20586900 |
| H  | 0.87428800 | 5.12879000 | -3.35567800 |
| H | -5.33608400 | -0.73476400 | -2.81607200 |
| H | -5.98619900 | 0.84050000 | -3.23750900 |
| H | -2.65745700 | 0.37307600 | -2.63496400 |
| H | -3.33946200 | 2.57571500 | 0.86823900 |
| H | -4.11524600 | -2.99495600 | 2.17637600 |
| H | -4.13657500 | -1.45843100 | 1.31888600 |
| H | -3.58545300 | -4.21744500 | 0.04069000 |
| H | -3.75694400 | -2.71634600 | -0.87707800 |
| H | -6.15439800 | -2.42448500 | -1.13800200 |
| H | -0.45617300 | 5.91587500 | 1.13242200 |
| H | -1.68634800 | 5.25472200 | 2.19418300 |
| H | 0.93223100 | -1.10576900 | 5.59627900 |
| H | -0.73021200 | -0.54724600 | 5.51174800 |
| H | 0.76860900 | -4.83138600 | 1.30494400 |
| H | 1.95022800 | -5.28547100 | 0.43399300 |
| H | 4.46145600 | -0.54212200 | -2.50036200 |
| H | 4.57648900 | 0.94561500 | -2.19183100 |
| H | 1.55435200 | -4.32574400 | -2.44651400 |
| H | 1.11884000 | -2.28861400 | -3.13452200 |
| H | 2.24161700 | -1.69373000 | -4.05271600 |
| C | 0.68682800 | 0.97986800 | 6.18120000 |
| H | 0.05232300 | 1.84466100 | 5.94059400 |
| H | 0.57087000 | 0.75423500 | 7.25323500 |
| H | 1.72687100 | 1.28157800 | 5.99790300 |
| C | 5.24232400 | -4.64827100 | 0.20869100 |
| H | 4.82020100 | -5.46620800 | -0.39389100 |
| H | 6.30060200 | -4.88378100 | 0.40207400 |
| H | 4.71520800 | -4.63543200 | 1.17282300 |
| C | 1.40001500 | 4.09262200 | -5.19766600 |
| H | 1.04537700 | 3.25343700 | -5.81383400 |
| H | 1.38001900 | 5.00571100 | -5.81437400 |
| H | 2.44161600 | 3.87398900 | -4.92615800 |
| C | 5.41911500 | 2.55998500 | 0.99417700 |
| H | 5.38292900 | 3.57607200 | 0.57043700 |
| H | 5.57316900 | 2.64640800 | 2.07712700 |
| H | 6.25439700 | 2.02371100 | 0.52517600 |
| C | 0.20684500 | 5.68558100 | 3.19770900 |
| H | -0.07745200 | 6.70752700 | 3.49588500 |
| H | 0.04620300 | 5.01770300 | 4.05568700 |
| H | 1.28296700 | 5.67309800 | 2.97798900 |
| C | -5.86751000 | -2.70987900 | 0.96132600 |
| H | -6.12511000 | -3.77651300 | 0.85351800 |
| H | -6.43653800 | -2.30701100 | 1.81040300 |
| C | -7.00950900 | -0.06859800 | -1.57622000 |
| H | -7.53961900 | 0.87295500 | -1.37566100 |
| H | -7.68154600 | -0.69465500 | -2.18604900 |
| N | -4.75749000 | 1.58048100 | -0.44138300 |
| H | -5.60332700 | 1.58907900 | 0.13003800 |

**$S_3$-state model in the peroxo state with O3-O6 bond at ~1.44Å**

**with Mn4-$H_2O$, -OH, charge -1. S=1 and His337-N$\delta$-H\* configuration**

\* Lack of the H-bond between the His337-N$\varepsilon$-H and O3 makes His337 drift away from our unrestrained cluster OEC model. Thus, His337 was removed from the model for this set of calculations.

| | | | |
|---|---:|---:|---:|
| C | 7.00698500 | -2.15054900 | 0.07852600 |
| C | 5.51216000 | -1.98845300 | 0.38396800 |
| O  |  4.85082900 | -3.01633400 |  0.65026900 |
| O  |  5.05902200 | -0.77552500 |  0.37430500 |
| C  | -2.86922200 | -0.77293400 |  4.99381200 |
| C  | -2.22251900 | -0.75574800 |  3.60704600 |
| O  | -1.18230400 | -1.39772700 |  3.38919200 |
| O  | -2.86331000 | -0.01219600 |  2.74786000 |
| C  | -0.54044500 |  6.43084100 | -0.18768000 |
| O  | -1.18747000 |  6.24133000 | -1.23156900 |
| C  | -1.34321700 |  5.32360700 |  2.00187300 |
| C  | -2.09010600 |  4.17377900 |  1.38178100 |
| C  | -2.38664100 |  2.92985500 |  1.89372800 |
| C  | -3.04343400 |  2.90812700 | -0.17587600 |
| N  | -2.96842500 |  2.14929000 |  0.91457300 |
| N  |  0.81633300 |  6.46810300 | -0.16675000 |
| C  |  1.85923100 |  4.59752600 | -1.47940300 |
| C  |  2.67481400 |  3.95992400 | -0.35145600 |
| C  |  2.55362000 |  2.42975300 | -0.24728300 |
| O  |  1.54782700 |  1.91998000 | -0.85032400 |
| O  |  3.41268700 |  1.83466700 |  0.45294800 |
| C  | -5.97872400 | -1.53944500 | -1.60487600 |
| C  | -4.58089800 | -1.16628600 | -1.10508300 |
| O  | -3.59240400 | -1.72768800 | -1.68338700 |
| O  | -4.52504000 | -0.33104400 | -0.15087300 |
| C  | -1.27740500 | -4.49517000 | -0.31784100 |
| O  | -0.77081900 | -4.21992000 |  0.78902600 |
| O  | -1.66408800 | -3.65491000 | -1.22783000 |
| Ca |  0.19702700 | -2.18188200 |  1.57304800 |
| O | -2.09591300 | -1.65853600 | 0.66691000 |
| O | 0.11835200 | -1.67933400 | -0.73150200 |
| O | -1.90607300 | 0.22610700 | -0.99189900 |
| O | 2.61452200 | -0.59043400 | -1.01138900 |
| O | 1.32005600 | -0.13801900 | 1.03184800 |
| O | -0.79971400 | 0.85446800 | -0.30708700 |
| Mn | -2.65269300 | 0.05562100 | 0.82594600 |
| Mn | -1.64006000 | -1.71658000 | -1.10499000 |
| Mn | 0.92999100 | 0.01204400 | -0.74595100 |
| Mn | 3.15183200 | -0.32471400 | 0.75610100 |
| O | 4.75048100 | 0.52051400 | -2.52004200 |
| O | 2.16402200 | -3.25638300 | 0.39410500 |
| O | 3.43864100 | -0.56752000 | 2.64606800 |
| O | 1.88453100 | -2.53727700 | 3.31578600 |
| C | -0.01130000 | -0.60607600 | -4.95190000 |
| C | -0.25263300 | -0.67182600 | -3.44288100 |
| O | -1.15876600 | -1.47847200 | -3.04089000 |
| O | 0.48156000 | 0.08680700 | -2.73995500 |
| H | 7.49562200 | -2.35948100 | 1.04720500 |
| H | 7.10980200 | -3.07500100 | -0.51012000 |
| H | -2.96797400 | 0.27583100 | 5.32081500 |
| H | -3.90320100 | -1.13255500 | 4.85977900 |
| H | -0.32415900 | 4.98115900 | 2.25242800 |
| H | -1.81478900 | 5.58374600 | 2.96384600 |
| H | -2.19008700 | 2.52705300 | 2.88139300 |
| H | -3.39780300 | 2.57795600 | -1.14592700 |
| H | 2.35733600 | 4.41481500 | -2.44522000 |
| H | 0.88165300 | 4.09934000 | -1.54379100 |
| H | 3.74501000 | 4.20837700 | -0.42243500 |
| H | 2.33655300 | 4.33286300 | 0.63312200 |
| H | 1.27308600 | 6.56216100 | 0.73537000 |
| H | -6.46543300 | -2.07382100 | -0.77063300 |
| H | -6.54205000 | -0.59924400 | -1.71810100 |
| H | 1.01292900 | -0.98398600 | -5.11017700 |
| H | 0.02765200 | 0.46149700 | -5.22075700 |
| H | 3.89428800 | 0.12834000 | -2.21529100 |
| H | 5.24413700 | 0.52982100 | -1.67948700 |
| H | 3.15656900 | -3.16082200 | 0.50749500 |
| H | 1.97571200 | -2.81624000 | -0.46202500 |
| H | 4.38031500 | -0.76142900 | 2.79642100 |
| H | 2.52712100 | -1.74286400 | 3.13836900 |
| H | 2.40433300 | -3.30648800 | 3.02314200 |
| C | -1.01643000 | -1.36740800 | -5.81422800 |
| H | -2.03564400 | -0.97949300 | -5.67520400 |
| H | -0.75533300 | -1.27669700 | -6.88039000 |
| H | -1.03631400 | -2.43354100 | -5.55083500 |
| C | 7.68225800 | -0.96837800 | -0.61715600 |
| H | 7.58545700 | -0.05085700 | -0.02024500 |
| H | 8.75459400 | -1.16832500 | -0.77311500 |
| H | 7.22577900 | -0.77642300 | -1.59923700 |
| C | -2.11735600 | -1.60851900 | 6.02742200 |
| H | -1.08795400 | -1.24698300 | 6.15598400 |
| H | -2.62695500 | -1.56910900 | 7.00283800 |
| H | -2.05164500 | -2.65914800 | 5.71289900 |
| C | -1.45761900 | -5.95662000 | -0.70795000 |
| H | -1.73099600 | -6.54532900 | 0.17736800 |
| H | -2.20557100 | -6.07331100 | -1.50136600 |
| H | -0.49073400 | -6.33253600 | -1.07710000 |
| C | -6.01824800 | -2.37651100 | -2.88236500 |
| H | -7.05762700 | -2.61858100 | -3.15482800 |
| H | -5.55797700 | -1.83958200 | -3.72381000 |
| H | -5.46240900 | -3.31502500 | -2.75569200 |
| C | 1.64243900 | 6.11015400 | -1.32893200 |
| H | 2.59902300 | 6.64740200 | -1.22707800 |
| H | 1.12492400 | 6.50316700 | -2.21516500 |
| C | -1.24306900 | 6.61622800 | 1.16044300 |
| H | -2.24590600 | 7.01083100 | 0.94201600 |
| H | -0.71375300 | 7.37470200 | 1.75982100 |
| N | -2.53764600 | 4.13959000 | 0.06896400 |
| H | -2.35319700 | 4.88159800 | -0.62311400 |

References

1. Becke, A.D., *DENSITY-FUNCTIONAL EXCHANGE-ENERGY APPROXIMATION WITH CORRECT ASYMPTOTIC-BEHAVIOR*. Physical Review A, 1988. **38**(6): p. 3098-3100.
2. Perdew, J.P. and W. Yue, *Accurate and simple density functional for the electronic exchange energy: Generalized gradient approximation*. Physical Review B, 1986. **33**(12): p. 8800-8802.
3. Weigend, F. and R. Ahlrichs, *Balanced basis sets of split valence, triple zeta valence and quadruple zeta valence quality for H to Rn: Design and assessment of accuracy*. Physical Chemistry Chemical Physics, 2005. **7**(18): p. 3297-3305.
4. Pushkar, Y., K.M. Davis, and M. Palenik, *Model of the Oxygen Evolving Complex Which is Highly Predisposed to O–O Bond Formation*. Journal of Physical Chemistry Letters, 2018. **9**: p. 3524-3531.
5. Pushkar, Y., et al., *Early Binding of Substrate Oxygen Is Responsible for a Spectroscopically Distinct S2 State in Photosystem II*. The Journal of Physical Chemistry Letters, 2019. **10**(17): p. 5284-5291.

6. Bury, G. and Y. Pushkar, *Insights from Ca2+→Sr2+ substitution on the mechanism of O-O bond formation in photosystem II*. Photosynthesis Research, 2024. **162**(2): p. 331-351.
7. Davis, K.M., et al., *Rapid Evolution of the Photosystem II Electronic Structure During Water Splitting*. 2015: p. https://arxiv.org/abs/1506.08862 (accessed May 17, 2017).
8. Davis, K.M. and Y.N. Pushkar, *Structure of the Oxygen Evolving Complex of Photosystem II at Room Temperature*. Journal of Physical Chemistry B, 2015. **119**(8): p. 3492-3498.
9. Davis, K.M., et al., *X-ray Emission Spectroscopy of Mn Coordination Complexes Toward Interpreting the Electronic Structure of the Oxygen-Evolving Complex of Photosystem II*. The Journal of Physical Chemistry C, 2016. **120**(6): p. 3326-3333.
10. Jensen, S.C., et al., *X-ray Emission Spectroscopy of Biomimetic Mn Coordination Complexes*. Journal of Physical Chemistry Letters, 2017. **8**(12): p. 2584-2589.
11. Siegbahn, P.E.M., *O-O bond formation in the S4 state of the oxygen-evolving complex in photosystem II*. Chemistry-a European Journal, 2006. **12**(36): p. 9217-9227.